# Interaction of discrete breathers with primary lattice defects in bcc Fe

Dmitry A. Terentyev[1], Andrii V. Dubinko[1], Volodymyr I. Dubinko[2], Sergey V. Dmitriev[3,4], Evgeny E. Zhurkin[4] and Mikhail V. Sorokin[5]

[1] SCK•CEN, Nuclear Materials Science Institute, Boeretang 200, Mol, 2400, Belgium
[2] NSC Kharkov Institute of Physics and Technology, Kharkov 61108, Ukraine
[3] Institute for Metals Superplasticity Problems RAS, Khalturin St. 39, Ufa 450001, Russia
[4] Tomsk State University, 36 Lenin Ave, Tomsk, 634050 Russia
[5] National Research Centre `Kurchatov Institute`, Kurchatov Square 1, 123182 Moscow, Russia

**Abstract** – The interaction of discrete breathers with the primary lattice defects in transition metals such as vacancy, dislocation, and surface is analyzed on the example of bcc iron employing atomistic simulations. Scattering of discrete breathers on the lattice defects induces localized atomic excitations, with intensity and relaxation time depending on the defect structure and breather kinetic energy. The dissipation of the intrinsic breather energy due to the scattering is computed and analyzed. It is concluded that the breather-to-defect energy transfer may stipulate the activation of the lattice defects causing unexpected athermal effects such as enhanced mass transfer or electroplasticity, already experimentally reported but so far not fully understood at the atomic-scale level.

**Introduction.** – Discrete breathers (DBs), a.k.a. intrinsic localized modes, are spatially localized large-amplitude vibrational modes in lattices that exhibit strong anharmonicity. They have been identified as the exact solutions to a number of model nonlinear systems possessing translational symmetry [1-5] and successfully observed experimentally in various physical systems [6]. It should be pointed out that experimental detection of DBs in crystals is a difficult task [7] and theoretical as well as computer simulation methods play an important role in investigating their properties.

An overwhelming majority of theoretical studies on DB deal with idealized one- or two-dimensional nonlinear lattices of coupled oscillators interacting via oversimplified pairwise potentials [4,8]. Studies of the DBs in three-dimensional space using realistic interatomic potentials were scarce and restricted to ionic crystals with NaCl structure, e.g., NaI [9-11], in graphane [12], in strained graphene [13,14], in some metals [15] and semiconductors [16]. Recently, DFT approach, that takes into account electronic structure of solids, was used for the first time to prove the existence of DBs in a crystalline solid considering graphane as an example [17].

Important for applications concept of quasi-breathers has been developed by Chechin with co-authors [18]. Spatially localized modes studied in the present work also should be termed quasi-breathers because they are not the exact solutions to the equations of atomic motion, but they have sufficiently long lifetime to participate in various processes and to influence the physical properties of the crystal.

DBs in real crystals radiate energy very slowly because their main vibrational frequency lies outside the phonon band. This happens because DB frequency is amplitude dependent and it increases (decreases) with amplitude for the hard (soft) type anharmonicity bifurcating up from the upper edge of phonon spectrum (down from the upper edge of the phonon gap). Thus, the soft-type anharmonicity DBs can exist only in the crystals having a gap in the phonon spectrum. Crystals with simple structure do not possess a gap in the phonon spectrum. It has been shown theoretically that the 1D chains with atoms interacting via classical pairwise potentials cannot support DB with hard-type anharmonicity [19]. However it can be demonstrated that introduction of the on-site potential in that model makes the existence of DB with frequency above the phonon spectrum possible by suppressing the dc displacements of the atoms (analog of the lattice thermal expansion due to the asymmetry of anharmonic interatomic potentials) and increasing the contribution of the hard core of the potential into atomic dynamics. In 2D crystal with Morse interatomic interactions hard-type anharmonicity DB are possible [20] because the close-packed atomic row, in which the DB is excited, experiences the action of the effective on-site potential induced by the rest of the crystal.

[a] E-mail: `dterenty@sckcen.be`
[b] Present address: SCK-CEN, Boeretang 200, Mol, 2400, Belgium

Oscillation frequencies of the DBs in ionic crystals [9,10] and in graphane and strained graphene [11-14] are found to be in the gaps of the phonon spectrum. However, transition metals and semiconductors possess no gap in phonon spectrum and thus DBs may exist only if their frequency is positioned above the phonon spectrum [15,16]. MD simulations based on realistic many-body interatomic potentials have proven that movable high-frequency DBs exist in fcc Ni as well as in bcc Nb and Fe [15,21,22]. Such DBs may exist in metals through the suppression of the dc displacements of the atoms by the effective on-site potential [20].

The threshold DB energy in metals is relatively small (of order of 1 eV) as compared to the formation energy of a stable Frenkel pair (several eV) [15,21,22]. Since DB in metals are highly mobile, they can efficiently transfer a concentrated vibrational energy over large distances along close-packed crystallographic directions [22], according to the MD simulations based on the potential derived from the embedded atom method [23]. Recently, a theoretical background has been proposed to ascribe the interaction of mobile DBs with lattice defects to rationalize the anomalous accelerated diffusion observed under conditions imposing intensive formation of DBs [24].

Only recently the interest of researchers has shifted to the study of the impact of DBs on the physical properties of materials [17,24-26]. This filed of research is principally new, and it lies at the conjunction of the nonlinear physics and material science, hence requiring input from the both disciplines. Evidence provided by the atomistic simulations about the existence of stable and mobile DBs in metals raises important questions: how DBs interact with primary lattice defects and how this interaction affects material properties. Certain technological processes (exposure to radiation, oscillating magnetic field, temperature and stress gradients, high-density electric current) as well as diagnostic procedures employing beams of energetic particles (e.g. electron microscopy, medical therapy) may cause continuous generation of DBs inside materials due to external lattice excitation, thus pumping a material with DB gas [23]. Depending on the interaction of DBs with natural defects, the presence of the DB gas may provoke, e.g., accelerated self-diffusion (DB-vacancy interaction), superplasticity and accelerated creep (interaction with dislocations), enhanced recrystallization (interaction with grain boundaries), dissolution or growth of secondary phase particles (interaction with precipitates).

Whether the release of intrinsic and kinetic DB energy by scattering on natural lattice defects such as vacancies, dislocations or grain boundaries would essentially influence materials properties or not is to be studied. If indeed, the scattering of a DB results in a localized atomic excitation, equivalent to a heating of material 'in a spot', this interaction should amplify the reaction rates due to effective reduction of the underlying activation barriers, as was recently proposed in [26-28]. Correspondingly, a qualitative and quantitative assessment of the interaction of DBs with lattice defects is not limited to academic interest and should be done to evaluate the possible impact of the DB gas on the material properties. So far, the interaction of moving DBs was considered only for vacancies and only in 1D [8] and 2D [20] crystals. Here, for the first time, we consider the interaction of moving DBs with lattice defects in bcc Fe using large scale MD simulations in 3D space. With the aid of modern *ab initio*-derived many body interatomic potentials (IAP), including 'magnetic' potential for bcc Fe, we investigate the interaction of DBs with a vacancy, dislocation and free surface.

**MD setup.** – The interaction of moving DBs with lattice defects is studied using classical MD simulations for 3D model of bcc Fe. To ensure that the obtained results are independent of the choice of the interatomic potential, we exploit two well-known IAPs including the *ab initio* - derived model by Chamati et al. [23] (used earlier in [21] to demonstrate the existence of a stable DB) and the 'magnetic' potential for bcc Fe developed by Dudarev *et al.* [28,29]. Although being semi-empirical, such IAPs are derived to account for the electronic charge distribution depending on the local atomic arrangement and are known to provide a good compromise between computationally expensive *ab initio* calculations and over-simplified pairwise potentials. Both IAPs have been widely and successfully used to model bulk and surface properties of bcc Fe as well as to study point-, extended- and interface-like lattice defects (see e.g. [30,31]).

MD simulations are done for the computational cell with three principal axes $x$, $y$ and $z$ oriented along the <111>, <-12-1> and <-101> crystallographic directions, respectively. The size of the MD supercell is chosen to be 100×3, 10×6, and 30×6 atomic planes along $x$, $y$ and $z$, respectively. Thus, the cell volume is 25×4×12 nm$^3$ and it contains about 140000 atoms. A DB is excited in the crystal by providing initial displacements along $x$ direction to six neighbouring atoms of a close-packed atomic row selected in the cell centre following the procedure proposed by Hizhnyakov et al. [21, 22]. The key feature of the procedure is the initial displacement of the two central atoms from their equilibrium position, which should oscillate in the anti-phase mode with respect to each other thus forming a stable DB. The central offset displacement will be referred in the following as $d_0$. Its variation determines the DB amplitude and frequency and, ultimately, its lifespan.

The movement of the DB is initiated by providing a momentum to the two central atoms forming the DB. The direction of the momentum coincides with the $x$-axis. Initial velocities of all atoms except for the two central atoms composing the DB are equal to zero.

The initial conditions used to excite moving DB are not exact and a part of the energy initially given to the crystal is radiated in the form of small-amplitude waves. Recall that only six atoms in the computational cell of 1.4×10$^5$ atoms are initially excited meaning that the energy density of phonons is four orders of magnitude smaller than in the core of DB and no noticeable effect of radiation on the mobility and propagation velocity of DBs was detected in the test runs with different computational cell sizes.

Periodic boundary conditions are applied along the $y$ and $z$ directions to model the interaction of a DB with a free surface, and 3D periodicity is imposed to study the interaction with a vacancy and with a ½<111>{110} edge dislocation dipole. In the latter case, an edge dislocation dipole is introduced in the crystal using a model of the periodic array of dislocations. Initially, the DB is introduced at a distance of (20-40)×$b$ from a defect of interest. Here, $b$ is the lattice spacing along the [111] direction, which is also the Burgers vector for the studied dislocation, and it is equal to $\sqrt{3}/2a_0$, where $a_0$ is the equilibrium bcc lattice constant, which is 2.86 Å according to the applied potentials [23,28]. Prior to initiating a DB in the crystal containing a defect, a full relaxation is achieved by the conjugate gradient method.

The use of periodic boundary conditions implies that actually not a single DB-vacancy interaction event is considered but this event happens simultaneously in a periodic array of cells. It is well-known that the type of boundary conditions can affect the properties of DBs [32]. For sufficiently large computational cell the interaction of the DB-vacancy system with its periodic images is small. It was found that in fcc Fe about 5 atoms in a close-packed atomic row constitute the core of a DB vibrating with large amplitudes. The same applies to the dc displacements of the atoms near the core of a DB. Spatial localization of the dc displacements is due to the effective on-site potential induced by the atoms surrounding the closed-packed atomic row hosting the DB. Test numerical runs have shown that further increase in the computational cell size does not noticeably affect the properties of standing and moving DBs.

In order to quantify motion of the DB and its interaction with the lattice defects, the atomic positions along the corresponding close-packed atomic row, and the deviations of potential energy of atoms from the initial values ($dE_P$) are analyzed. The time evolution of the atomic displacements and $dE_P$ is used to find the DB frequency, the transferred kinetic energy, the duration of the interaction with a defect (i.e., the period during which the atoms in the vicinity of a defect are excited) and the energy dissipated to phonons due to the DB-defect scattering. Through the text, we present the results obtained using the IAP from Ref. [28], which has been derived accounting for the properties of point defects in bcc Fe as derived with the help of *ab initio* methods. However, the comparison with the other IAP is provided to demonstrate the robustness of the obtained results and discussed implications.

**MD results.** – Recall that the DB was initiated by the initial displacement of two neighbouring atoms by ±$d_0$ along <111> crystallographic direction. A parametric study to identify the range of $d_0$ providing a stable standing DB revealed that the most stable DBs with the lifespan from 3.5 to 4 ps were produced with $d_0$ from 0.26 to 0.35 Å ($d_0/a_0$ from 0.092 to 0.122) as demonstrated in Fig. 1, the left ordinate. DB lifetime at the edges of their existence range is equal to just a few oscillation periods. Note that the standing DB frequency (the right ordinate of Fig. 1) ranges from 10 to 14 THz and it is close to the Debye frequency of bcc Fe and it grows with increasing $d_0$ as expected from the hard type anharmonicity of the considered vibrational mode. It can be seen from Fig. 1 that there is a certain range of $d_0$ resulting in a stable DB. Application of a displacement $d_0$ smaller than 0.26 Å does not provide enough potential energy for the two oscillators to initiate a stable DB and the atomic oscillations decay quickly, while a displacement larger than 0.45 Å ($d_0/a_0$ larger than 0.157) generates a chain of focusons [33,34]. We note that focusing collisions (focusons) [33,34] are produced in the recoil events and transfer energy along close packed directions of the lattice, but there is no interstitial transport by a focusing collision. In an ideal lattice a focuson travels in a close packed direction and loses its energy continuously by small portions in each lattice site, which determines its propagation range as a function of initial energy.

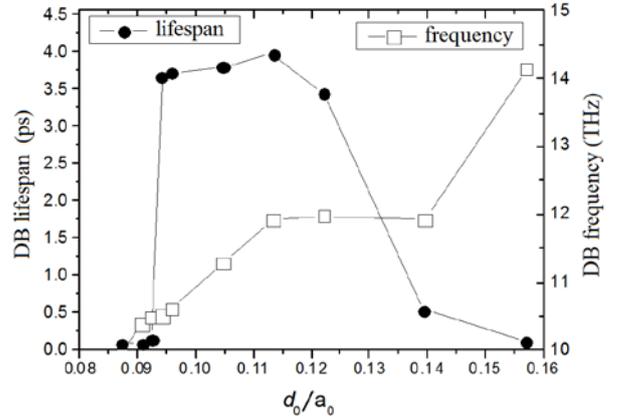

Fig. 1: Lifespan and frequency of standing DB as the function of the relative initial displacement $d_0/a_0$ for the IAP derived in [28].

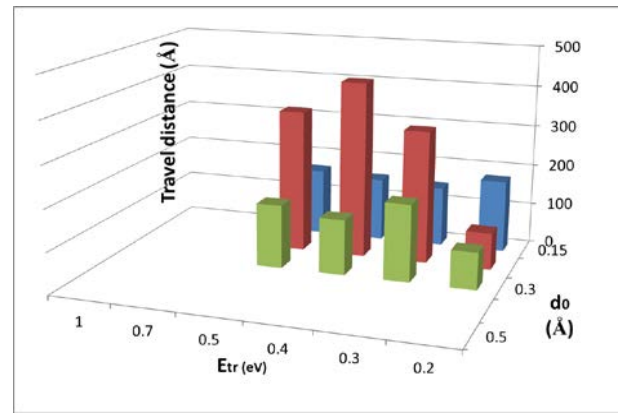

Fig. 2: Travel distance of mobile DBs as a function of initial displacement $d_0$ and translational energy $E_{tr}$ for the IAP derived in [23].

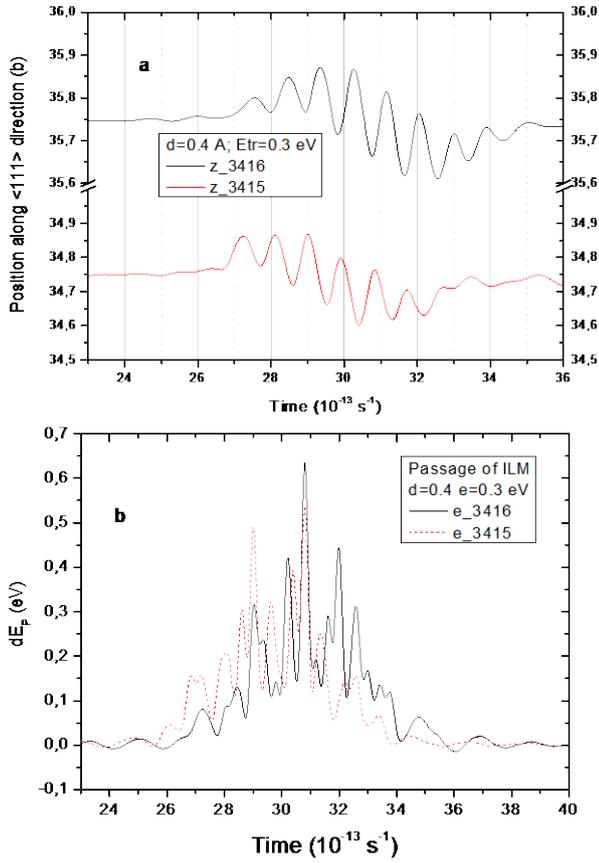

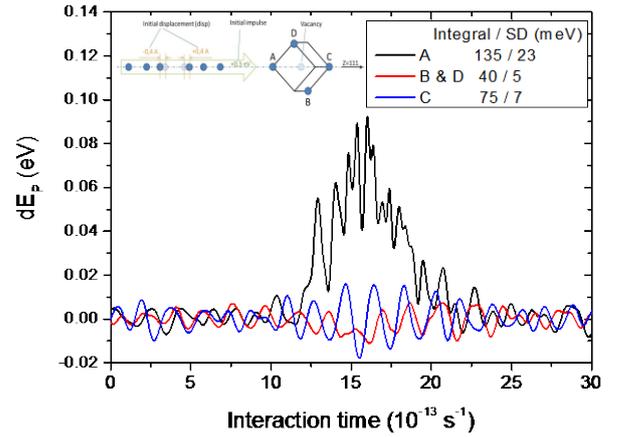

Fig. 3: (a) *x* coordinate as the function of time for two neighbouring atoms, 3415 and 3416, in the close-packed row hosting DB during the passage of DB excited with $d_0 = 0.4$ Å, $E_{tr}= 0.3$ eV. (b) Deviation of the potential energy of the atoms from the ground state during the passage of DB. The dc displacements of atoms is about five times smaller than their maximal vibration amplitude and they cannot be clearly seen in the figure.

Potential energy of the crystal after the introduction of the initial displacements $\pm d_0$ for the two neighboring atoms can be used as an estimation of the DB energy. The increase of the total potential energy of the system due to the introduction of a DB with a minimum displacement $d_0 = 0.27$Å is 2.96 eV, which can be viewed as a threshold energy of the DB formation. This energy subsequently partitions between intrinsic kinetic and potential energy of the atoms constituting DB. Even though standing DB has an intrinsic kinetic energy (due to atomic oscillations), it does not move because its translational kinetic energy along the *x*-axis is zero.

The DB can be set in motion by providing translational kinetic energy ($E_{tr}$) to the two central DB atoms along the *x*-axis.

Fig. 4: Deviation of the potential energy of the atoms surrounding a vacancy from the ground state caused by the scattering with the DB ($d_0 = 0.4$ Å, $E_{tr}= 0.3$ eV). The inset provides schematics of the position of inspected atoms with respect to the vacancy and impinging DB.

The DB travel distance was determined by inspecting the evolution of the atomic displacements in the <111> row in which the DB was created. Raw simulation data can be found in the supplementary materials. The file Moving_DB.avi shows a DB moving along close-packed atomic row. The files Slow_DB_hits_vacancy.avi and Fast_DB_hits_vacancy.avi contrast the results of collision of relatively slow and relatively fast DB with a vacancy.

The travel range of DBs as a function of $d_0$ and $E_{tr}$ is presented in Fig.2 for the IAP from ref. [23]. It appears that mobile DBs can exist in the interval of initial parameters: $d_0$ from 0.15 to 0.5Å ($E_{tr}$ from 0.2 to 0.6 eV). DB velocity ranges from 300 to 2000 m/s while travel distances ranges from several dozens to several hundreds of atomic spaces depending on the $d_0$, $E_{tr}$ and utilized IAP. Note that the DB traveled distance has a maximum as a function of the DB kinetic energy, $E_{tr}$. This can be explained by the fact that lifespan has also maximum and the longer the DB lives the further it propagates.

Passage of DB that exhibits rather long travel distance (at least 60***b***) through a pair of neighbouring atoms arbitrary selected in the close-packed row hosting the DB is presented in Fig.3. In (a) *x* coordinates of the atoms and in (b) potential energy of the atoms are shown as the functions of time. Fig.3a shows that the atom with index 3415 starts to move at $t_1$=2.7 ps, while the next atom at $t_2$=2.75 ps. The two atoms vibrate almost out-of-phase for about 1ps (~10 oscillations) and then oscillations cease. The DB moves with a speed of 2.14 km/s, which is about the half speed of sound in bcc Fe. The translational kinetic energy of the DB is about 0.54 eV, which is shared among two core atoms, with the energy of 0.27 eV per atom. This is close to the initial kinetic energy $E_{tr}$=0.3 eV given to the atoms to initiate the DB movement. It can be seen from Fig.3b that the amplitude of the potential energy change for an atom can reach almost 1 eV. In a simplified 'thermodynamic' analogy, a moving DB can be viewed as an atom-size spot heated above $10^4$ K propagating though the crystal at sub-sonic speed. The potential or kinetic energy of individual atoms in the core of DB averaged over period reaches up to 0.5 eV, which corresponds to the effective temperature ~5000 K, however, as a matter of fact, standing or moving DB does not produce any topological defects or any other signs of local lattice melting.

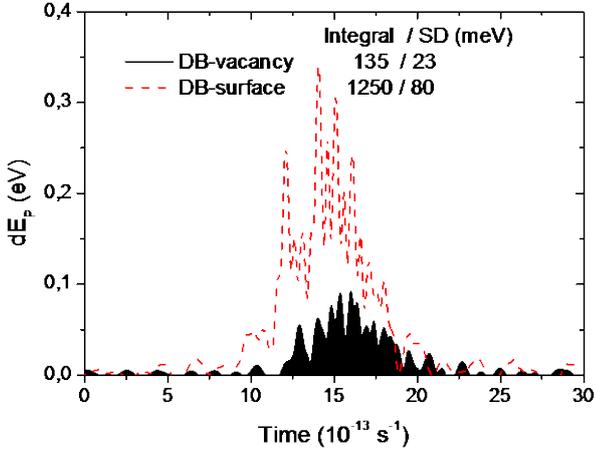

Fig. 5: Comparison of the atom energy deviation from the ground state for the cases of DB-vacancy and DB-surface interactions. DB excitation parameters $d_0 = 0.4$ Å, $E_{tr}=0.3$ eV.

*DB-vacancy interaction.* In the defect-free lattice, the dissipation of kinetic energy of the moving DB is small which is why it can cover a long distance. However, the defects break translational lattice symmetry and therefore are expected to act as scattering centres for the DBs. As the first example, the scattering of DBs with different $E_{tr}$ was studied in a crystal containing a vacancy. For relatively fast DBs (large $d_0$ and $E_{tr}$ values resulting in the DB velocity exciding 1 km/s) the interaction time and transferred energy were found to be essentially the same. In most cases, we observed that DBs were reflected from a vacancy, moved away from it (in the opposite direction) for several lattice units and faded out. Following the oscillations of atoms surrounding a vacancy, a typical excitation time (or DB-vacancy interaction time) was found to be about 3 ps (~30 atomic oscillations).

The excitation of atoms surrounding a vacancy is best presented by the variation of their potential energy, given in Fig.4. The inset figure provides a schematic of the atomic location and the DB propagation direction. One can see that the strongest excitation occurs for the atom A experiencing the front collision with the DB. Other atoms acquire much weaker oscillations and consequently the variation of the potential energy. The integral under the curve for the frontal atom and a standard deviation (SD) of its potential energy from the ground state far exceeds that for other adjacent atoms (see the text inset of Fig. 4).

Relatively slow DBs (small $d_0$ and $E_{tr}$ values resulting in the DB velocity < 500 m/s) are reflected from the vacancy elastically: they turn back at about 5*b* from the vacancy without touching it, while the atoms surrounding the vacancy remain essentially unperturbed by the DB reflection. The total propagation range and lifespan of the DB is not changed by the collision, which confirms its elastic nature.

Parameters of the DB interaction with the vacancy are listed in Table 1.

*DB-surface interaction.* Interaction of a fast DB with a free surface with {111} orientation, i.e. normal to the DB propagation direction, results in excitation of essentially one surface atom and reflection of the DB back to the bulk. The reflected DB decays rather quickly in about two-three lattice periods. Apparently, the interaction of a fast DB with free surface results in essential energy loss. The oscillation of the potential energy of the surface atom hit by the DB is shown in Fig.5, and the curve obtained for the single vacancy (frontal atom A) is reproduced for comparison. The energy oscillation of the surface atom is essentially higher than that for the near-vacancy atom. Such a strong excitation of the surface atom is apparently related to the break-up of the lattice symmetry at the free surface. This results in the loss of a significant part of DB energy and leads to its fast decay. The integral under the curve corresponds to the total energy $E_{tot}$ (listed in Table 1) transferred to the defect in the process of DB-defect interaction.

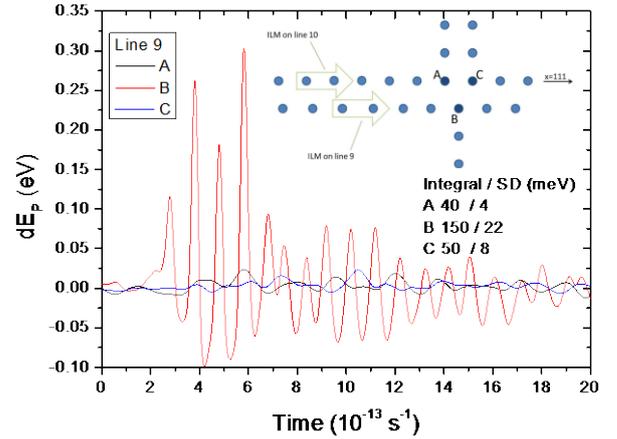

Fig. 6: Variation of the potential energy of atoms surrounding the core of ½<111>{110} edge dislocation caused by the scattering of the DB ($d_0 = 0.4$ Å, $E_{tr}=0.3$ eV) moving along the atomic row toward atom B.

Slow DBs are reflected from the surface almost elastically: their total propagation range and lifespan are not changed by the collision. At the same time, the surface atoms are displaced from equilibrium positions but the oscillation of their potential energy is much smaller than in the case of the interaction with a fast DB (see Table 1).

*DB-dislocation interaction.* The scattering of DBs on the core of the edge dislocation was studied for different parameters. The DB passed through the dislocation in all the cases, except for the direct hit in the lower part of the dislocation core, as shown in the inset schematics in Fig.6. The only intensive DB scattering was registered for the DB moving along the atomic row toward atom B. The excitation of the three atoms forming the dislocation core is presented for this case in Fig.6. Apparently, only the frontally hit atom B exhibits essential vibration. The integral under the curve for the atom B is 0.15 eV, which is comparable with that for the near-vacancy atom A. The dislocation excitation time is about 2 ps, which is close to the excitation time for DB-vacancy or DB-free surface interaction.

**Discussion.** – Scattering of DBs on the lattice defects, investigated here by MD simulations, resulted in the localized atomic excitations causing essential perturbation of potential and kinetic energy of the atoms surrounding the defects. Within the relaxation time of this excitations (listed in Table 1), it is natural to expect the amplification of the reaction rates involving defects subjected to the DB scattering [24,26,27]. For example, the scattering of the DB on a vacancy will raise its probability to perform a migration jump, scattering on a dislocation - facilitate the kink nucleation or unpinning from an obstacle, scattering on a void - facilitate the atom evaporation or the vacancy emission inside the crystal, etc.

Table 1: Parameters of the DB interaction with the lattice defects: excitation time τ, excitation amplitude $E_m$, total transferred energy $E_{tot}$, and standard deviation $\langle E \rangle_{SD}$.

| Defect / number of atoms involved | τ (ps) | $E_m$ (eV) | $E_{tot}$ (eV) | $\langle E \rangle_{SD}$ (eV) |
|---|---|---|---|---|
| Data obtained with IAP [28] | | | | |
| Vacancy / 8 | 3 | 0.1 | 0.37 | 0.045 |
| Free surface, fast DB / 1 | 3 | 0.35 | 0.135 | 0.08 |
| Free surface, slow DB / 1 | 2 | 0.35 | 0.12 | 0.09 |
| Dislocation core / 3 | 2 | 0.3 | 0.24 | 0.03 |
| Data obtained with IAP [23] | | | | |
| Vacancy / 8 | 3 | 0.02 | 0.49 | 0.008 |
| Free surface, fast DB / 1 | 3 | 0.12 | 0.077 | 0.03 |
| Free surface, slow DB / 1 | 4 | 0.06 | 0.058 | 0.013 |
| Dislocation core / 3 | 2.5 | 0.2 | 0.03 | 0.048 |

Table 1 summarizes the main characteristics of the defect neighbourhood relaxation as a result of the interaction with the DB. $\langle E \rangle_{SD}$ is the standard deviation of the potential energy of atoms surrounding the defect from the initial level calculated over the excitation time τ. The number of atoms in the volume activated by DB is also specified in Table 1. The results are presented for the two IAPs.

To better underline the significance of the contribution of the DB-defect interactions to the crystal structure transformations let us analyse the related enhancement of an arbitrary thermally activated process. One possibility is to consider the amplification mechanism based on modification of the classical Kramers escape rate from a potential well due to a periodic modulation of the well depth (i.e. the reaction barrier height), which is an archetype model for chemical reactions since 1940 [26]. The reaction rate averaged over a time span exceeding the modulation period has been shown to increase exponentially with the ratio of the modulation amplitude $E_m$ and temperature:

$$\langle \dot{R}_K \rangle = \dot{R}_K I_0 \left( \frac{E_m}{k_b T} \right), \quad (1)$$

$$\dot{R}_K = \omega \exp \left( -\frac{E_a}{k_B T} \right), \quad (2)$$

where ω and $E_a$ are the attempt frequency factor (of order of atomic oscillation period) and the activation energy, respectively, $k_b$ is the Boltzmann constant and $T$ is the temperature. The amplification factor $I_0(x)$ is the zero order modified Bessel function of the first kind, which does not depend on the modulation frequency or on the mean barrier height. Even though the periodic forcing due to DB scattering is too weak to induce a direct athermal reaction (if $E_m < E_a$), it amplifies the average reaction rate drastically given that $E_m/k_b T >> 1$ condition is respected [26]. Since the excitation lifetime is longer than the typical atomic oscillation period (τ>>1/ω), one may consider the barrier modulation amplitude to be constant within relaxation of the defect after its interaction with the DB. By taking the excitation amplitudes reported in Table 1, one reveals that $E_m/ k_b T >>1$ realizes in the case of fast DBs for $T$ up to 600K, i.e., well within the technological range of applications. On the other hand, slow DBs may affect lattice defects less strongly but they are not destroyed by the scattering on natural defects. Therefore, these will build up the DB gas resulting in small stochastic modulations of the potential barriers, which has been shown to enhance the reaction rates via effective reduction of the underlying reaction barriers [24,27] as:

$$\langle \dot{R} \rangle = \omega \exp \left( -E_a^{DB}/k_b T \right), \quad (3)$$

$$E_a^{DB} = E_a - \frac{\langle E \rangle_{SD}^2}{2k_b T}, \quad (4)$$

where $\langle E \rangle_{SD}$ is the standard deviation of the potential energy of atoms surrounding the defect, evaluated in the present study (Table 1). The physical reactions, which will be the most affected by the background DB gas, include diffusion of the radiation-induced lattice defects (point defects and 1D glissile dislocation loops), break-up of vacancy-solute and pure vacancy clusters, nucleation of kinks on dislocations and kink diffusion, rearrangement of grain boundary core atoms and grain boundary sliding, and other atomic-scale activation events requiring the energy of the order of the vacancy migration barrier (i.e. ~0.5 eV). Another way to reconcile the effect of DB gas on the enhancement of thermally activated processes is to assign in to the raise of lattice vibration entropy as a result of continuous scattering of slow DBs on defects.

**Conclusions.** – To summarize, we have analyzed stability, mobility and interaction of discrete breathers with defects of crystal lattice in bcc Fe using atomistic simulations in 3D space. The most typical defects in metals such as vacancy, dislocation and free surface were considered. It was shown that two scenarios of interaction can be observed depending on the DB velocity. Relatively fast DB were expected to essentially modulate the frequency of thermally activated events within a relatively narrow relaxation time; while slow DB accumulate in the form of DB gas, whose presence can be viewed as increase of lattice vibration entropy. Both types of the interactions will enhance the reaction rate of thermally activated atomic-level processes, as is shown on the basis of the analysis applying the modified Kramers escape rate theory. The results obtained with the use of two different interatomic potentials differ quantitatively but not qualitatively.

With respect to the technological applications of Fe as well as Fe-based alloys and steels, the above demonstrated and discussed breather-defect interactions should result in the modification of such important properties as ductility, creep, swelling, phase separation, electroplasticity, heterogamous segregation and others diffusion-limited processes responsible for degradation of mechanical properties under ageing and/or irradiation [24,26,27,35-37].

Among open problems let us mention the effect of temperature on the DB properties [38].

*Acknowledgments*. S.V.D. gratefully acknowledges financial support from the Tomsk State University Academic D.I. Mendeleev Fund Program and the Russian Science Foundation grant N 14-13-00982.